\documentclass[referee]{svjour3}
\usepackage{graphicx}
\usepackage{rotating}
\usepackage{amssymb}
\usepackage{mathptmx}
\usepackage[numbers]{natbib}
\makeatletter
\journalname{}

\newcommand{\figref}[1]{Fig.~\ref{#1}}

\newcommand{\eqref}[1]{Eq.~\ref{#1}}

\newcommand\jcap{J. Cos. Astro. Phys. }
\newcommand\araa{Ann. Rev. Astron. Astrophys. }
\newcommand\procspie{SPIE Conf. Series }
\newcommand\aap{Astron. Astrophys. }
\newcommand\apjs{Astrophys. J., Suppl. Ser. }
\newcommand\jltp{J. Low Temp. Phys. }
\newcommand\jai{J. Astro. Inst. }
\newcommand\sst{Sup. Sci. Tech. }
\newcommand\prl{Phys. Rev. Lett. }

\newcommand\apj{Astrophys. J. }
\newcommand\apl{Appl. Phys. Lett. }
\newcommand\aanda{Astron. Astrophys. }
\newcommand\ao{Appl. Opt. }
\newcommand\arcmp{Annu. Rev. Condens. Matter Phys. }

\begin{document}

\newcommand{\hdblarrow}{H\makebox[0.9ex][l]{$\downdownarrows$}-}
\title{Low Temperature Detectors for CMB Imaging Arrays}

\author{J. Hubmayr \and 
J.E.~Austermann \and 
J.A.~Beall \and 
D.T.~Becker \and
B.~Dober \and
S.M.~Duff \and 
J.~Gao \and
G.C.~Hilton \and
C.M.~McKenney \and
J.N.~Ullom \and
J.~Van~Lanen \and
M.R.~Vissers}
\institute{Quantum Sensors Group, National Institute of Standards and Technology\\ Boulder, CO 80305, USA\\Tel.:303.497.6164\\
\email{hubmayr@nist.gov}}

\maketitle

\begin{abstract}
We review advances in low temperature detector (LTD) arrays for Cosmic Microwave Background (CMB) polarization experiments, with a particular emphasis on imaging arrays.  
We briefly motivate the science case, which has spurred a large number of independent experimental efforts.  
We describe the challenges associated with CMB polarization measurements and how these challenges impact LTD design.
Key aspects of an ideal CMB polarization imaging array are developed and compared to the current state-of-the-art.  
These aspects include dual-polarization-sensitivity, background-limited detection over a 10:1 bandwidth ratio, and frequency independent angular responses.  
Although existing technology lacks all of this capability, today's CMB imaging arrays achieve many of these ideals and are highly advanced superconducting integrated circuits.  
Deployed arrays map the sky with pixels that contain elements for beam formation, polarization diplexing, passband definition in multiple frequency channels, and bolometric sensing.
Several detector architectures are presented.
We comment on the implementation of both transition-edge-sensor bolometers and microwave kinetic inductance detectors for CMB applications.  
Lastly, we discuss fabrication capability in the context of next-generation instruments that call for $\sim 10^6$ sensors.


\end{abstract}

\section{Introduction}

Observational cosmology using the Cosmic Microwave Background (CMB) is an extremely active field.  
Over 15 sub-orbital experiments are in various stages of the design, deployment, and data analysis project lifecycle.  
Final analysis from the Planck \cite{collaboration2014planck} satellite is nearing completion, and the JAXA-led LiteBIRD satellite is under development \cite{matsumura2014mission}.  
This enthusiasm exists for two main reasons.  
First, the science reach is broad, touching on different disciplines such as cosmology, astrophysics, and particle physics.  
CMB measurements explore the physics of the extremely early universe, constrain the properties of neutrinos, and enable multiple probes of the growth rate of structure.    
For overviews on CMB science see \cite{challinor2012cmb,s4science,kk2016,finelli2018inflation,divalentino2018cosmo,challinor2018lensing,melin2018cluster}.
Second, detection of the gravitational wave signature as a curl component in CMB polarization on degree angular scales, referred to as primordial B-mode polarization, is the most promising technique to determine what produced the initial conditions of standard big bang cosmology \cite{seljak1997measuring, kamionkowski1997probe, seljak1997signature}.  
Furthermore, through this endeavor we may craft an experimental probe of quantum gravity and investigate physics at grand unification energy scales.  
These broad science goals demand instrument capability of sufficient dynamic range in angular resolution and sensitivity such that multiple experiments exist in order to address them.  

In terms of detector development, the trend has been to include more capability on-chip.  
Modern CMB detector arrays not only sense sky power but also provide elements for beam formation, polarization diplexing, and passband definition.
CMB polarimeters are indeed superconducting integrated circuits (ICs).  
The vast majority of experiments are imaging arrays, which utilize polarization-sensitive arrays of voltage-biased transition-edge-sensor (TES) bolometers \cite{irwin2005transition}.    
However interest in microwave kinetic inductance detectors (MKIDs) \cite{day2003,baselmans2012kinetic} for CMB applications is increasing.
   
In this review, we first state the challenges associated with CMB polarization measurements that impact detector design.  
We then describe an ideal imaging focal plane and show that modern detector arrays achieve many of these ideals.  
We discuss the trade-offs between TES bolometers and MKIDs when implementing for CMB polarimetry.  
Lastly, we detail wafer production capabilities in the context of next-generation CMB experiments that demand very large detector counts, such as the Simons Observatory \cite{so-web}, BICEP array \cite{grayson2016bicep3}, ALI-CPT \cite{li2017tibets}, and CMB-Stage-IV \cite{s4science,s4instrument}.  
 
\section{Measurement Challenges}

\begin{figure}[t]
\begin{center}
\includegraphics[width=1.0\linewidth, keepaspectratio]{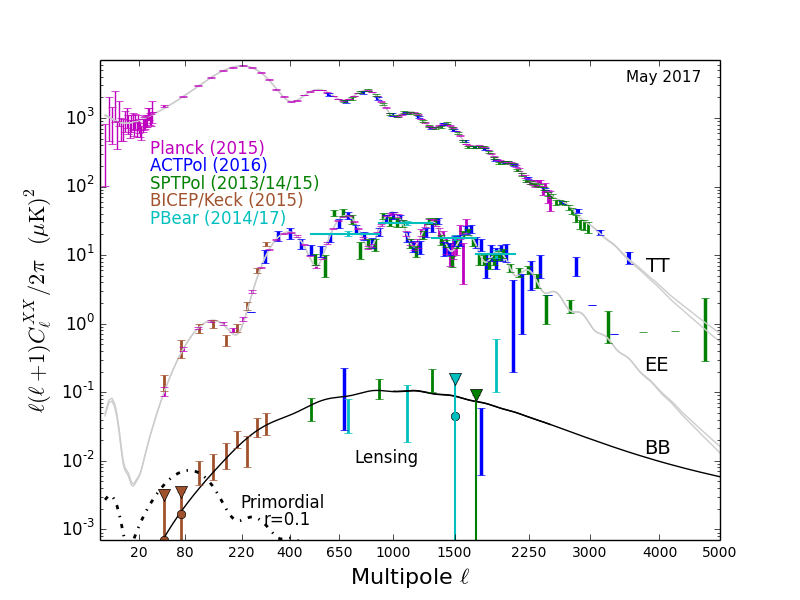}
\caption{The CMB angular power spectra (temperature anisotropy, E-mode polarization, and B-mode polarization are labeled  `TT', 'EE', and 'BB' respectively) with state-of-the-art measurements as of May 2017 from 
\cite{planck2015ps, louis2017atacama, keisler2015measurements, crites2015measurements, hanson2013detection, bk-vi, ade2014measurement, ade2017measurement}.  
Cosmological B-mode power consists of two components: primordial gravitational waves ( labeled `primordial' and parameterized by the tensor-to-scalar ratio $r$) and gravitational lensing of E-mode into B-mode (labeled `lensing').  
Figure courtesy L. Page. (Color figure online.)
\label{fig:ps}}
\end{center}
\end{figure}
Precision measurements of CMB polarization are challenging for several reasons.
Foremost, the signal-to-noise ratio is extremely low.    
Figure~\ref{fig:ps} shows the CMB angular power spectra together with state-of-the-art measurements.  
The amplitude of primordial gravitational waves is parameterized by the tensor-to-scalar ratio $r$.  
The determination of, or constraints on $r$ distinguish between the various models of inflation.  
For reference, the plot shows $r$~=0.1, consistent with current upper limits \cite{bk-vi}.
Thus even in the most favorable scenario, the detection of primordial B-modes requires the measurement of $\sim$~80~nK fluctuations on top of the 2.73~K uniform background.  
Consequently, instruments must not only have extremely high sensitivity, which has motivated the use of large arrays of sensors, but must also mitigate sources of systematic error.       
Instrumental polarization, which undesirably converts power from the unpolarized CMB temperature anisotropy into polarization, is of particular concern because the temperature anisotropy is $>1500 \times$ stronger than the most stringent upper limit on the amplitude of primordial B-modes at $\ell = 80$.  
Thus, 0.1\% temperature-to-polarization leakage manifests a false signal that is larger than the expected cosmological signal. 
More discussion of CMB systematic errors can be found in \cite{odea2007systematic, shimon2008cmb}.

The 2.73~K CMB blackbody spectrum dictates measurement at millimeter (mm) wavelengths, which presents unique challenges.  
In mm-wave optical systems, diffraction is substantial and requires the use of quasioptical methods \cite{goldsmith1998quasioptical} in both design and system evaluation.  
Signal attenuation arises due to beam divergence and loss both in on-wafer transmission lines and in camera optical elements, as it is challenging to find materials with low loss tangent in the mm regime (see Ref.~\cite{lamb1996miscellaneous} for example).  
Additionally there is a dearth of mm-wave measurement tools, which increases the development time of LTDs.    
For example, a system capable of cryogenic scattering-parameter (S-parameter) measurements at $\sim$150~GHz would allow device builders to characterize individual components of their integrated circuit (IC).   
Current practices, limited by available testing infrastructure, rely on characterizing the full IC.  
Subsequent attribution of anomalous behavior to one of the multiple circuit components can therefore be challenging.  

\begin{figure}[t]
\begin{center}
\includegraphics[width=0.70\linewidth, keepaspectratio]{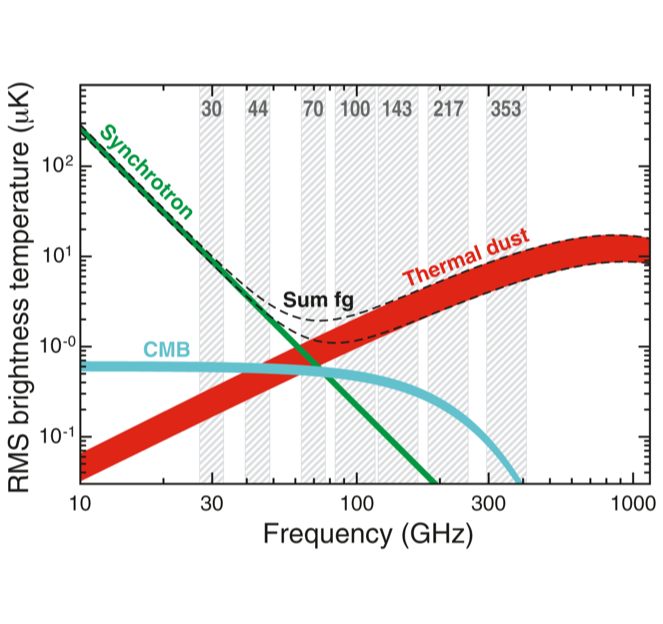}
\caption{RMS brightness temperature of CMB E-mode polarization and the main foreground sources: synchrotron and thermal dust, for sky fractions 73\% and 93\% (lower and upper dashed curves).  
Note that the CMB B-mode brightness temperature is at least an order of magnitude below the curve labeled `CMB', illustrating the seriousness of the foreground mitigation challenge.  
For a version of this figure which includes B-modes at various levels see Ref.~\cite{remazeilles2018comp}.  
The presence of high amplitude foregrounds motivates observations at multiple frequencies for component separation. 
Figure reproduced from \cite{planckX} (Color figure online.)
\label{fig:foreground}}
\end{center}
\end{figure}
Of course, the CMB is not the only source of emission at mm-wavelengths.  
The atmosphere is highly emissive at particular wavelengths and attenuates the signal to a certain extent at all wavelengths.  
Ground-based instruments therefore observe from dry sites such as the Atacama Desert and the South Pole through discrete atmospheric transmission windows.
However, the greater challenge is that astrophysical foregrounds (synchrotron emission, galactic dust, anomalous microwave emission (AME), free-free, etc.) also emit at mm-wavelengths.
Figure~\ref{fig:foreground} shows that the amplitude of polarized foreground sources is large relative to CMB polarization.  
To separate these sources from the CMB, we rely on the fact that the frequency spectrum of foreground sources differs from a thermal source.  
Broad frequency coverage ($\sim$30-300~GHz) is therefore necessary for component separation. 

In addition, many science drivers require $\sim$~arcminute resolution, necessitating a $>$5~m class telescope.    
The need for high angular resolution impacts detectors in two main ways.  
First, large telescopes with high throughput create large focal plane areas that require many detector wafers to fill.  
This increases wafer volume, which is a key challenge for the field and is discussed in Sec.~\ref{sec:waferproduction}.  
Second, the relative expense of the telescope motivates the instrument builder to collect as many photons as possible.  
In this regard broad bandwidth detection is advantageous and has driven the development of multichroic detector architectures discussed in Sec.~\ref{sec:ideal}.

These five aspects of the measurement (low signal-to-noise, mm-wavelengths, systematic errors, foregrounds, and high angular resolution) drive CMB instrument and detector design.
In the next section, we describe an ideal CMB imaging array.

\section{Ideal CMB Imaging Focal Plane}
\label{sec:ideal}

\begin{figure}[t]
\begin{center}
\includegraphics[width=1.0\linewidth, keepaspectratio]{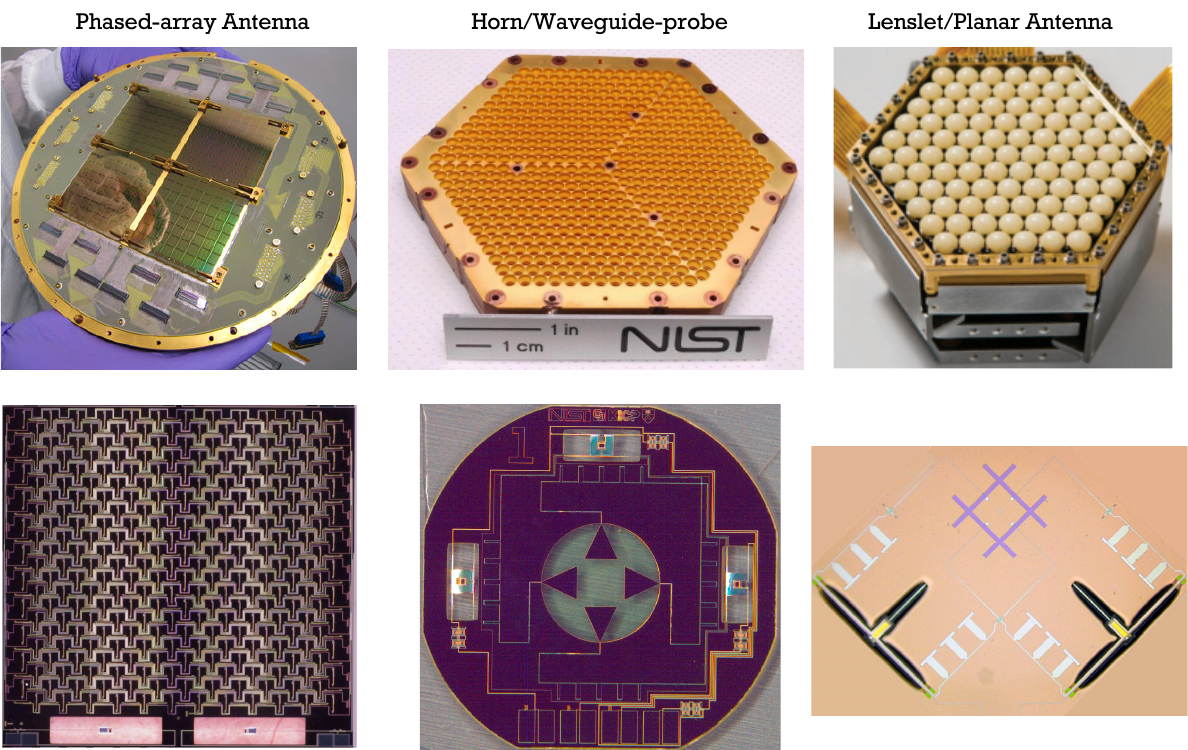}
\caption{Superconducting ICs used for the CMB power spectrum measurements presented in \figref{fig:ps}.  
The top row shows the coupling technology and the bottom shows the pixel IC.  
{\it Left:} Phased antenna arrays developed at Caltech/JPL \cite{kuo2009antenna,obrient2012antenna}.  
In this architecture the beam is formed on-chip by the phased-array, and thus a flat antireflection wafer serves as the only off-detector-wafer focal plane optical coupling component. 
{\it Center:} Feedhorn/waveguide-probe-coupled detectors using silicon feedhorns (pictured) have been developed at NIST \cite{yoon2009feedhorn}.  
An alternative probe-coupled detector design (not pictured) has been developed at NASA/Goddard \cite{chuss2012electromagnetic}.    
{\it Right:} Lenslet/planar antenna arrays have been developed at UC, Berkeley \cite{arnold2012bolometric}.
In this case, the antenna gain is increased with an extended hemispherical lenslet.   
(Color figure online.)
\label{fig:cmbdet}}
\end{center}
\end{figure}
An ideal CMB imaging array would map the sky quickly with polarization sensitive detectors in multiple frequency bands over a 10:1 bandwidth ratio ($\sim$30-300~GHz).  
Despite diffraction, this array would couple efficiently to receiver optics over the entire bandwidth.  
 
Mapping the CMB is most efficient with the use of photon-noise-limited detectors--that is detectors for which the dominant noise source arises from the incident photons and not from the detector or other sources.
The photon noise equivalent power ($NEP$), defined as the uncertainty in detected power in a bandwidth of 1~Hz, is given by \cite{zmuidzinas2003thermal}
\begin{equation}
\label{eqn:photonnep}
	NEP^2_{photon} =  \int_0^\infty 2 \frac{dP}{d\nu} h \nu \Big(1+\eta(\nu)m(\nu)\Big) d\nu.
\end{equation}
The first term describes photon shot noise whereas the second term is a correction to Poisson statistics due to wave bunching \cite{boyd1982photon}.  
$\nu$ is the band averaged center frequency, $\frac{dP}{d\nu}$ is the absorbed power per unit frequency, $\eta(\nu)$ is optical efficiency, and $m(\nu)$ is the photon occupation number, which for a thermal source $m(\nu)~=~(\mathrm{exp}[h\nu / k_b T]-1)^{-1}$.  

As photon noise is irreducible, even a noiseless detector maps the sky at a very slow rate.    
To illustrate the point, it is instructive to ask the question: `using a single noiseless detector, how long would it take to discover primordial gravitational waves?'  
The observation time $t$ required to achieve a polarization map depth $M$ in a sky area $A_{sky}$ may be estimated from  
\begin{equation}
	\label{eq:ms}
	t = 2\big(\frac{NET_{cmb}}{M}\big)^2 \frac{A_{sky}}{\eta_o},
\end{equation}
where $NET_{cmb}$ is the noise equivalent temperature relative to the CMB (the $rms$ noise in the detector after 1 second of observing the 2.73K background) and $\eta_o$ is the observation efficiency (which should not be confused with optical efficiency $\eta$).        
In recent work Errard et al. show that a 5$\sigma$ detection of $r=0.07$, which is the current best upper limit \cite{bk-vi}, requires 1\% of the sky to be mapped in polarization to a depth $M$~=~7$\mu$K-arcmin \cite{errard2016robust}.  
Assuming perfect observing efficiency ($\eta_o$~=~1) and a single, noiseless detector with $NET_{cmb} \sim 10~\mu K \sqrt{s}$ 
\footnote{This calculation assumes that photon noise from the CMB is the only noise source and that the noiseless detector perfectly couples to a single electromagnetic mode over the frequency range 125~GHz--165~GHz.}, the required observation time is only two months.  

Unfortunately for the CMB experimentalist, this estimation is overly optimistic in a number of ways.  
Due to diffraction and photon absorption in commonly used materials, the optical efficiency of real mm-wave systems is $\eta < 0.5$.  
Other sources of emission (atmosphere, optical components, cryostat walls, etc) load the detector and produce a photon noise level that can be larger than the photon noise generated by the CMB.  
Additionally the observation efficiency can never be unity.   
Ground-based sites only have favorable observing conditions for approximately half the year; 
turn around time in raster scans are cut; 
detectors are inoperable during mK refrigerator cycles; 
calibration source observations require time; 
and large volumes of data are cut in order to avoid systematic errors.  
For a ground-based instrument, these factors lead to $\eta_o \sim 0.2$, and assuming an excellent sensitivity $NET_{cmb} = 200 \mu\mathrm{K}\sqrt{s}$, \eqref{eq:ms} yields a sobering result: 384 years.  
Space-based instruments fare better but still require 5 years of observation ($\eta_o$ near unity is possible from space and $NET_{\mathrm{cmb}}~=~50 \mu \mathrm{K}\sqrt{\mathrm{s}}$ has been demonstrated in the Planck 143~GHz channels \cite{planck2011first}).  
For a discussion on the relative sensitivity of ground versus space-based CMB instruments see appendix A of Ref.~\cite{delabrouille2018exploring}. 

\begin{figure}[t]
\begin{center}
\includegraphics[width=1.0\linewidth, keepaspectratio]{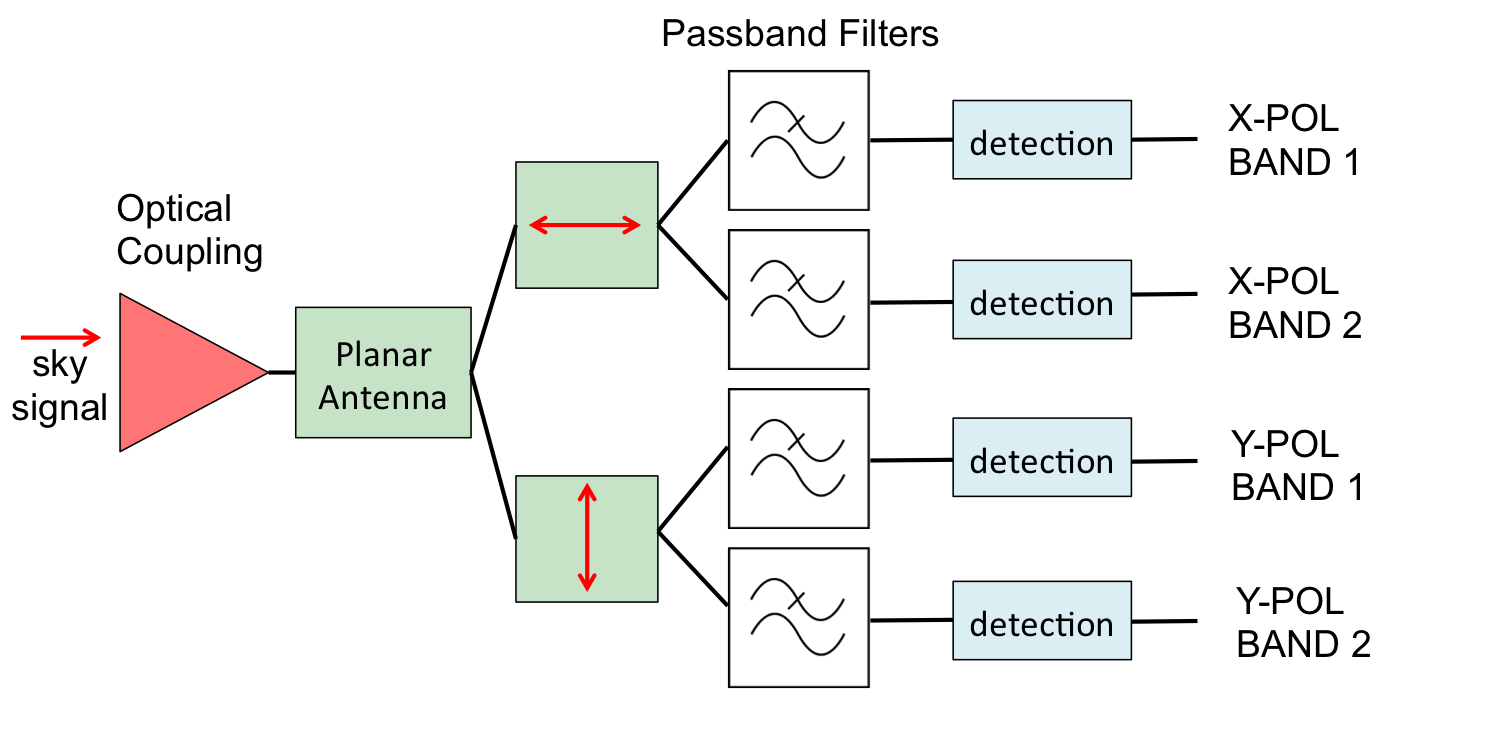}
\caption{Conceptual diagram of a dual-polarization-sensitive, multichroic detector.  (Color figure online.)
\label{fig:mc-schematic}}
\end{center}
\end{figure}

This simplistic calculation illustrates a main point that strongly impacts detector technology: in order to efficiently map CMB polarization, we must collect more light.   
This light may be collected in focal planes with either physically larger detector pixels if high angular resolution is not required (i.e. SWIPE \cite{gualtieri2016multi} and the spectrometer PIXIE \cite{nagler2016multimode}) or by producing arrays of detectors that are each sensitive to a single spatial mode, so-called `single-mode-coupled' detectors.    
Several types of single-mode-coupled, dual-polarization-sensitive detector arrays have been developed and are distinguished largely by optical coupling approach.  
These superconducting ICs for CMB polarization measurements (summarized in \figref{fig:cmbdet}) have several attributes of the ideal CMB polarization imaging array.  
The sensors have proven to be near photon-noise-limited \cite{grace2014actpol,thornton2016actpol,ade2014bicep2}.
Each is dual-polarization sensitive, which avoids otherwise reflecting half the photons, and the observation passband is defined with on-chip filtering.  
High technical maturity of these architectures is evidenced by their use in ACTPol \cite{thornton2016actpol}, SPTpol \cite{austermann2012sptpol}, BICEP2/Keck Array \cite{bk5}, and PolarBear \cite{arnold2010polarbear} to produce the power spectrum measurements shown in \figref{fig:ps}.  

These arrays, however, fall short of the ideal CMB imaging array in terms of bandwidth.  
The fractional detection bandwidth is $\sim$~30\%, far from a 10:1 bandwidth ratio.  
Significant effort has gone into the development of multichroic detectors, shown schematically in \figref{fig:mc-schematic}, which couple broadband mm-wave radiation onto a superconducting transmission line and channelize into several frequency bands.  
Multichroic phased-arrays, feehorn-coupled, and lenslet-coupled detectors have all been demonstrated in the lab \cite{datta2014horn,hubmayr2015feedhorn,suzuki2012multi}.  

\begin{figure}[t]
\begin{center}
\includegraphics[width=1.0\linewidth, keepaspectratio]{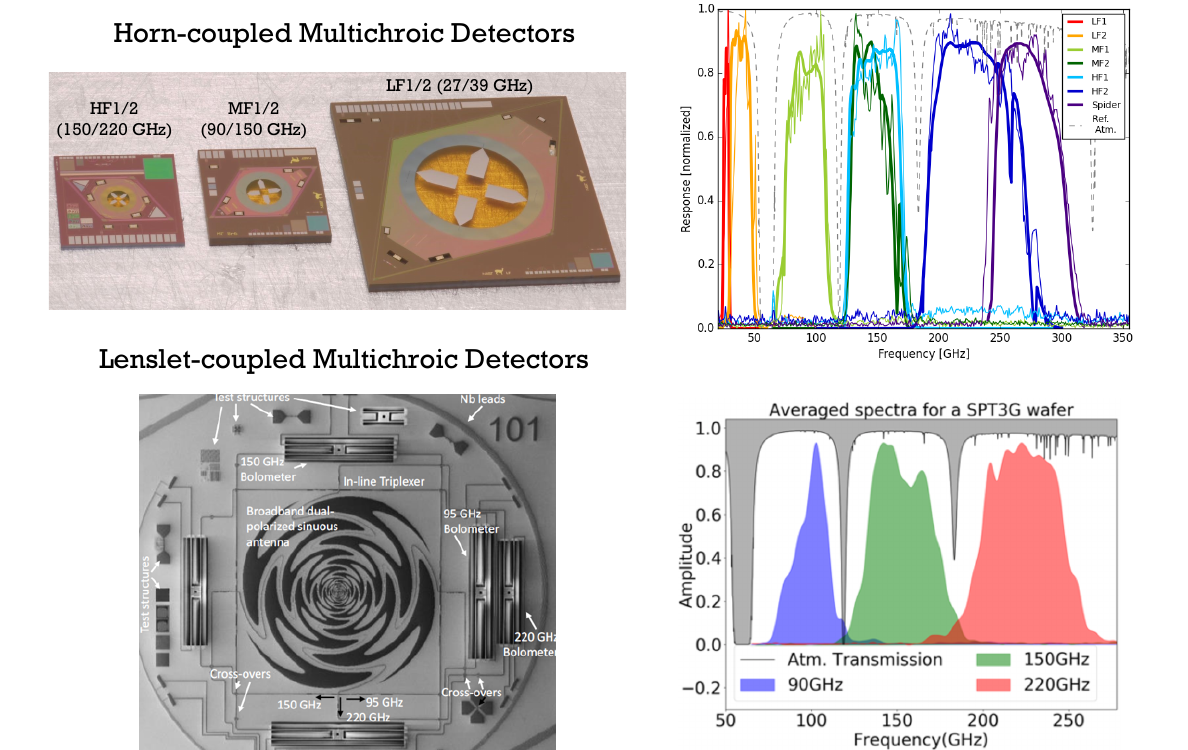}
\caption{Dual-polarization sensitive, multichroic detectors for CMB.  
{\it Top:} 2.3:1 bandwidth ratio multichroic, horn-coupled detectors have been developed at NIST \cite{mcmahon2012multichroic, datta2016design} in multiple frequency-scaled versions to span the 10:1 total bandwidth ratio.  
The {\it top right} figure is a compilation of the simulated (thick lines) and measured (thin lines) passbands for the three frequency-scaled dichroic pixels (pictured) and a single frequency band SPIDER 280 GHz pixel \cite{hubmayr2016design} (not pictured).  
For reference, the dashed-grey line shows a model of atmospheric transmission.
{\it Bottom:} 3:1 bandwidth ratio sinuous antenna, which couple to lenslets, have been developed at UC, Berkeley \cite{suzuki2012multi,westbrook2016development} and ANL \cite{posada2015fabrication}.  
This photograph show a device fabricated at ANL.  
The {\it bottom right} figure presents the averaged passbands of a wafer comprised of 90/150/220~GHz trichroic pixels.  
Figure reproduced with permissions from \cite{pan2017ltd} (Color figure online.)
\label{fig:mc}}
\end{center}
\end{figure}

\figref{fig:mc} shows two examples of deployed multichroic detectors. 
Feedhorn-coupled arrays with 2.3:1 bandwidth dichroic detectors at 90/150~GHz have been deployed in ACTPol \cite{thornton2016actpol,datta2016design}.
Second generation multichroic horn-coupled arrays at 90/150~GHz \cite{choi2017ltd,crowley2017ltd} and 150/220~GHz \cite{ho2017highly,crowley2016characterization} have been deployed in advanced ACTPol.  
Lenslet-coupled arrays with 3:1 bandwidth sinuous antennae coupled to trichroic 90/150/220~GHz detectors have been fielded in SPT-3G \cite{anderson2017ltd}, 
and 90/150~GHz as well as 150/220~GHz lenslet-coupled sinuous detectors are soon to be deployed in Simons Array \cite{arnold2014simons}. 
Although multichroic detectors are well-established, CMB data collected with such arrays have yet to be published.

To achieve the frequency coverage required for foreground characterization, the design of both single frequency and multichroic pixels has been frequency-scaled.    
The horn-coupled dichroic detectors in \figref{fig:mc} illustrate one example of this scaling.
Realizing a useful 10:1 bandwidth ratio in one pixel not only requires expanding the working frequency range of all components in the IC---a formidable challenge---but also requires a frequency independent beam size.  
The mapping speed optimization of an array is a trade-off between detector count and efficient coupling to receiver optics.  
Several discussions on the topic can be found in the literature \cite{gbg2002, s4instrument}. 
Currently implemented multichroic detectors use a fixed diffracting aperture, which produces frequency-dependent beam sizes and thus cannot efficiently couple to receiver optics at all frequencies.    
Datta et al. \cite{datta2014horn} argue that a properly sized 2:1 bandwidth pixel with dichroic detectors achieves 85\% of the optimal mapping speed for each frequency band.    
The outcome is increased spectral resolution and a 70\% mapping speed boost using the same focal plane footprint of a single frequency array.    
The addition of a third band, using 3:1 bandwidth ratio detectors of the same aperture size, achieves only 50\% of the mapping speed of an array optimized for that frequency\footnote{This calculation depends slightly on the Lyot stop temperature.}.  
Following this logic, one gains little by producing a fixed aperture 10:1 bandwidth pixel.

To abate the issue, one would choose an optimized pixel size per frequency band within a multichroic array.    
By use of hierarchical phased arrays of sinuous antennae, Cukierman et al. recently demonstrated near frequency-independent beam sizes in a single polarization at band centers near 90, 150, and 220~GHz  \cite{cukierman2017ltd}.  
This proof of principle demonstrates a viable path toward 10:1 bandwidth ratio CMB detectors.  
Several engineering challenges remain (signal routing topology, low-loss transmission lines, well-controlled beam systematics), and these topics are future areas of research for the technology.  

\section{Sensor Technology}

\begin{figure}[t]
\begin{center}
\includegraphics[width=1.0\linewidth, keepaspectratio]{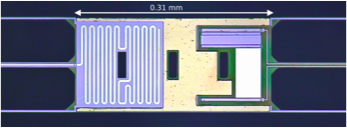}
\caption{Example transition-edge-sensor (TES) bolometer fabricated at JPL.  
Optical power is coupled onto a superconducting transmission line and dissipated on the thermally isolated membrane by use of the meandered gold structure on the left.  
This power is compensated by the electrical power dissipated in the voltage-biased TES located on the right.  
(Color figure online.)
\label{fig:tesbolo}}
\end{center}
\end{figure}
\begin{figure}[t]
\begin{center}
\includegraphics[width=.70\linewidth, keepaspectratio]{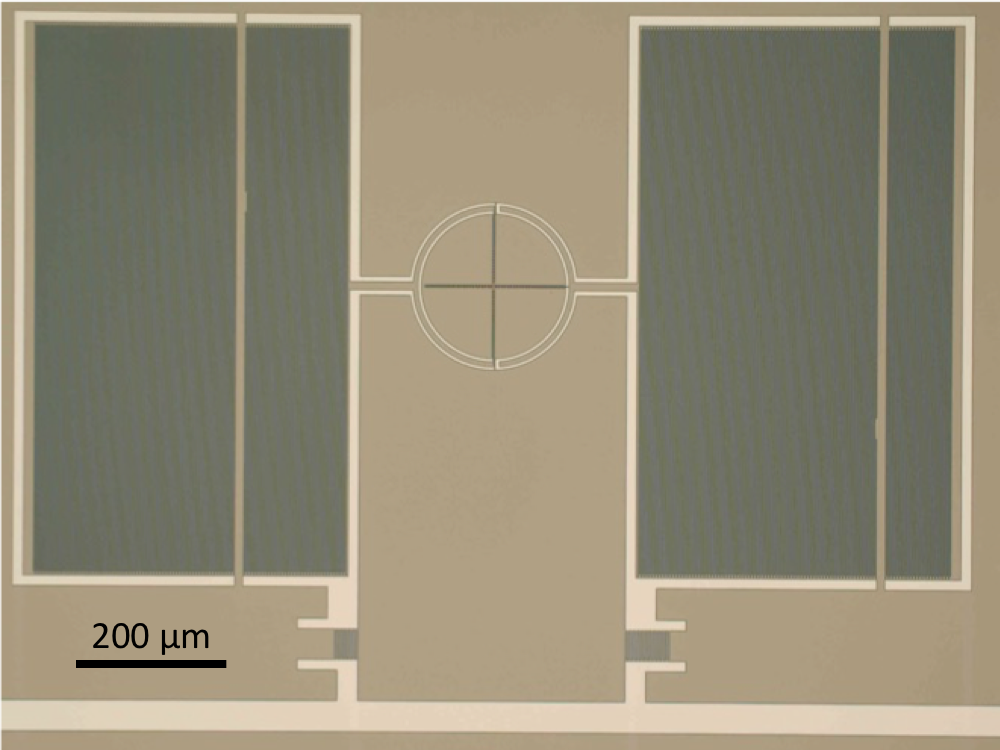}
\caption{BLAST-TNG lumped element MKIDs fabricated at NIST \cite{galitzki2014next}.  
The cross feature in the center contains two absorbers, one for each linear polarization, that also are the inductive element of an LC resonator.  
The inductors couple to interdigitated capacitors (large structures to the left/right of the center cross) that are in turn capacitively coupled to the microwave feedline at the bottom. 
(Color figure online.)
\label{fig:mkid}}
\end{center}
\end{figure}

The two leading sensor technology candidates for CMB polarization measurements are the voltage-biased transition-edge-sensor (TES) bolometer \cite{lee1998voltage,irwin2005transition} and the microwave kinetic inductance detector (MKID) \cite{day2003, baselmans2012kinetic, zmuidzinas2012a}.  
Examples of each are shown in \figref{fig:tesbolo} and \figref{fig:mkid}.  
Voltage-biased TES bolometers are thermal detectors that operate by the principle of electrical substitution.  
Joule power dissipated in the sensor directly compensates for changes in coupled optical power.  
Signals are read out by use of SQUID-based multiplexers.  
TES bolometers have been the workhorse sensor for CMB measurements in the last decade.  
In contrast, MKIDs have yet to be fielded in a CMB experiment but have been developed and deployed at relevant wavelengths \cite{golwala2012a, catalano2014performance, adam2018nika2}.  
MKIDs have several attractive features for application to CMB, and indeed research in this direction is active \cite{oguri2016groundbird, johnson2017development, tang2017fabrication, dominjon2017development}.
MKIDs are non-equilibrium pair-breaking devices, in which signal results from a change in the number of quasiparticles within a high quality factor (high-Q) superconducting resonator.  
A major strength of the approach is that high-Q resonators may be frequency division multiplexed in large numbers at either RF or microwave frequencies.  

The ultimate sensitivity of TES bolometers and MKIDs for CMB applications is comparable as other authors have presented \cite{lowitz2014comparison}.  
The fundamental detector noise of the TES bolometer is thermal fluctuation noise associated with phonon transport across the weak thermal link.  
The phonon noise $NEP$ may be expressed as \cite{mather1982bolometer,irwin2005transition} 
\begin{equation}
\label{eq:nep-g}
NEP_{G}^2~=~F_{link}4k_bT^2G,
\end{equation}
where $T$ is the bolometer temperature, $G$ is the thermal conductance, and $F_{link}(T,T_b)$ accounts for temperature gradients along the thermally isolating structure to the base temperature $T_b$.   

Generation-recombination noise is the fundamental detector noise source in MKIDs, and has been extensively studied \cite{devisser2011number,deVisser2012}.  
In the limit of photon dominated quasiparticle generation, the detector noise of the MKID is only due to quasiparticle recombination.  
Expressed as an $NEP$, the recombination noise is \cite{zmuidzinas2012a,flanigan2016photon} 
\begin{equation}
\label{eq:nep-gr}
NEP_r^2 = 4P\Delta/\eta_{pb}.
\end{equation}   
Here $P$ is the radiative photon load, $\Delta$ is the superconducting gap energy, and $\eta_{pb}$ is the efficiency of converting photons to quasiparticles.  
 
For both sensor types, the ratio of detector $NEP$ to photon $NEP$ is independent of optical load $P$.
Rather the detector mapping speed degradation $(NEP_{det} / NEP_{photon})^2 \sim T_c / \nu$, where $T_c$ is the superconducting transition temperature of the device.  
For TES bolometers, this quantity is determined by phonon transport properties including the material choice for thermal isolation, 
the ratio $T_b/T_c$, but is mostly influenced by the ratio of total power dissipation required to maintain the bolometer at $T_c$ (i.e. the bolometer saturation power) to the absorbed optical power.   
This `safety factor' is an engineering choice that must be larger than one for the bolometer to be cooled to $T_c$ and thus be operable.  
The detector $NEP$ scales as the square root of the safety factor.  
Common safety factor values are between 2 and 4, which for a $T_c~=~160$~mK bolometer observing in the 150~GHz band decreases the mapping speed by $\sim$~33\% and $\sim$~66\%, respectively.  

For MKIDs the exact degradation factor depends on the detailed quasiparticle dynamics in the resonator; 
however practical implementation for CMB detection may limit the value to near 100\%.     
The quasiparticle generation efficiency $\eta_{pb}$ in Eq.~\ref{eq:nep-gr} is a non-trivial function of the photon frequency to gap ratio--unity at the gap edge, scaling as 1/$\nu$ in the range $2\Delta < h\nu < 4\Delta$, and plateaus to a constant and material dependent number $\sim$~0.6 at $h\nu~>~4\Delta$ \cite{guruswamy2014quasiparticle}.  
This detailed picture has been experimentally verified in BCS superconductors \cite{devisser2015non}.  
Flanigan et al. \cite{flanigan2016photon} argue and demonstrably show that a MKID observing photons near the gap ($2\Delta < h\nu < 4\Delta$) has a recombination $NEP$ equal to the photon shot noise $NEP$.  
The authors note that the result is expected given the symmetry between uncorrelated pair-breaking events (photons) and uncorrelated pair-recombination events.    
Thus the detector mapping speed degradation of a MKID in this configuration is equal to the ratio of the photon shot noise $NEP$ to total photon $NEP$ given in Eq.~\ref{eqn:photonnep}, and this value is close to 1 --  
the MKID detector noise is nearly equal to the photon noise.  
For detection of CMB photons near the peak of the 2.73~K blackbody function, we are necessarily in the regime $2\Delta < h\nu < 4\Delta$ since the photon energy is low, and $\Delta$ cannot be made arbitrarily low because otherwise thermally generated quasiparticles degrade the $NEP$.  
Equation~\ref{eq:nep-gr} assumes $T_b \ll T_c$ to avoid thermally generated quasiparticle fluctuations that would otherwise degrade sensitivity.      

Avoiding NEP degradation from thermally generated quasiparticles is particularly challenging to meet for synchrotron monitoring channels at low observation frequencies.  
A background-limited MKID sensitive to 30~GHz requires $T_b \lessapprox 60$~mK.  
Note that the thermal MKID architecture, originally developed for x-ray applications \cite{quaranta2013xray, miceli2014towards} and now under development for CMB polarization measurements \cite{steinbach2018thermal}, would not have this constraint as the approach is bolometric.   

These discussions illustrate that in either case, reaching the fundamental sensitivity limit in a large fraction of sensors on an array is challenging 
and strongly depends on the details of not only the detector design but also the receiver configuration.   
Moreover, unexpected noise sources, such as TLS noise in MKIDs \cite{gao2008semiempirical} or `excess-noise' in TES bolometers \cite{irwin2005transition}, can be more limiting than fundamental noise sources.     

Fabrication of MKID arrays has long been touted as simple---in principle a single deposition and etch step.  
However, when MKIDs are coupled to antennas or waveguide probes, as is currently under development \cite{tang2017fabrication,johnson2016polarization,ji2014design} and desired for CMB applications, fabrication is no longer simple and fast.  
In these implementations, both TES-based and MKID-based approaches have fabrication challenges that are comparable, and the fabrication rates ought to be similar.  
From a detector fabrication standpoint, one offers no strong advantage over the other.
Rather the strength of MKIDs for CMB applications is two-fold.  
First, cryogenic readout requires only a broadband microwave LNA (a simplification from necessary TES biasing and SQUID multiplexing components).  
Second, detector packaging is vastly simplified since the number of interconnects to the array is reduced from thousands of DC connections to few microwave transmission lines.  
This simplification is expected to improve end-to-end yield and increase the development rate.    
  
\section{Wafer Production}
\label{sec:waferproduction}

The detector count required for upcoming experiments, and in particular ground-based experiments, demands a large increase in wafer fabrication capability.   
For example, CMB-Stage-IV requires $\sim~10^6$ detectors, whereas the total number of TES bolometers fielded in CMB polarization experiments to date is $\sim$~40,000.  
While this number is an order of magnitude below the aspirations of next-generation experiments, wafer production in the US CMB community is already impressive and expanding.
Through the projects SPIDER \cite{fraisse2013spider}, BICEP2, \cite{ade2014bicep2}, Keck Array \cite{staniszewski2012keck}, and BICEP3 \cite{ahmed2014bicep3} Caltech/JPL have alone produced and fielded 96 TES bolometer arrays, each fabricated on 100~mm diameter wafers.       
What is required for CMB-Stage-IV is perhaps a factor of three larger if the wafer size is increased.  
    
A common industry approach to increase production rates and decrease costs has been to fabricate devices on larger substrates.  
Major semiconductor foundries routinely fabricate on 300~mm diameter wafers and a consortium of large companies is exploring a move to 450~mm wafers.  
For LTD production, upfront tooling costs prohibit the use of such large wafers, and furthermore the LTD wafer volume---miniscule compared to the semiconductor industry---neither justifies nor requires the use of such large wafers.  
Most CMB wafer fabrication facilities have recently migrated to 150~mm wafers, the maximum expected wafer size of LTDs in the upcoming years.   
This move has increased the development rate by a factor between 2 and 4, since for a fixed number of sensors fewer parts need to be fabricated, assembled, and cryogenically tested.  
The first 150~mm wafer articles of feedhorn-coupled detectors \cite{ho2017highly,choi2017ltd} fabricated at NIST and lenslet/sinuous-coupled detectors \cite{anderson2017ltd} fabricated at Argonne National Laboratory have been deployed.  
Community-wide, $\sim$~100,000 sensors fabricated on 150~mm are planned for the near-term experiments Simons Array\cite{arnold2014simons}, Simons Observatory \cite{so-web}, Ali-CPT \cite{li2017probing}, and BICEP array \cite{hui2017ltd}.  

\begin{figure}[t]
\begin{minipage}{4.75in}
  \centering
  \raisebox{-0.5\height}{\includegraphics[width=0.45\linewidth, keepaspectratio]{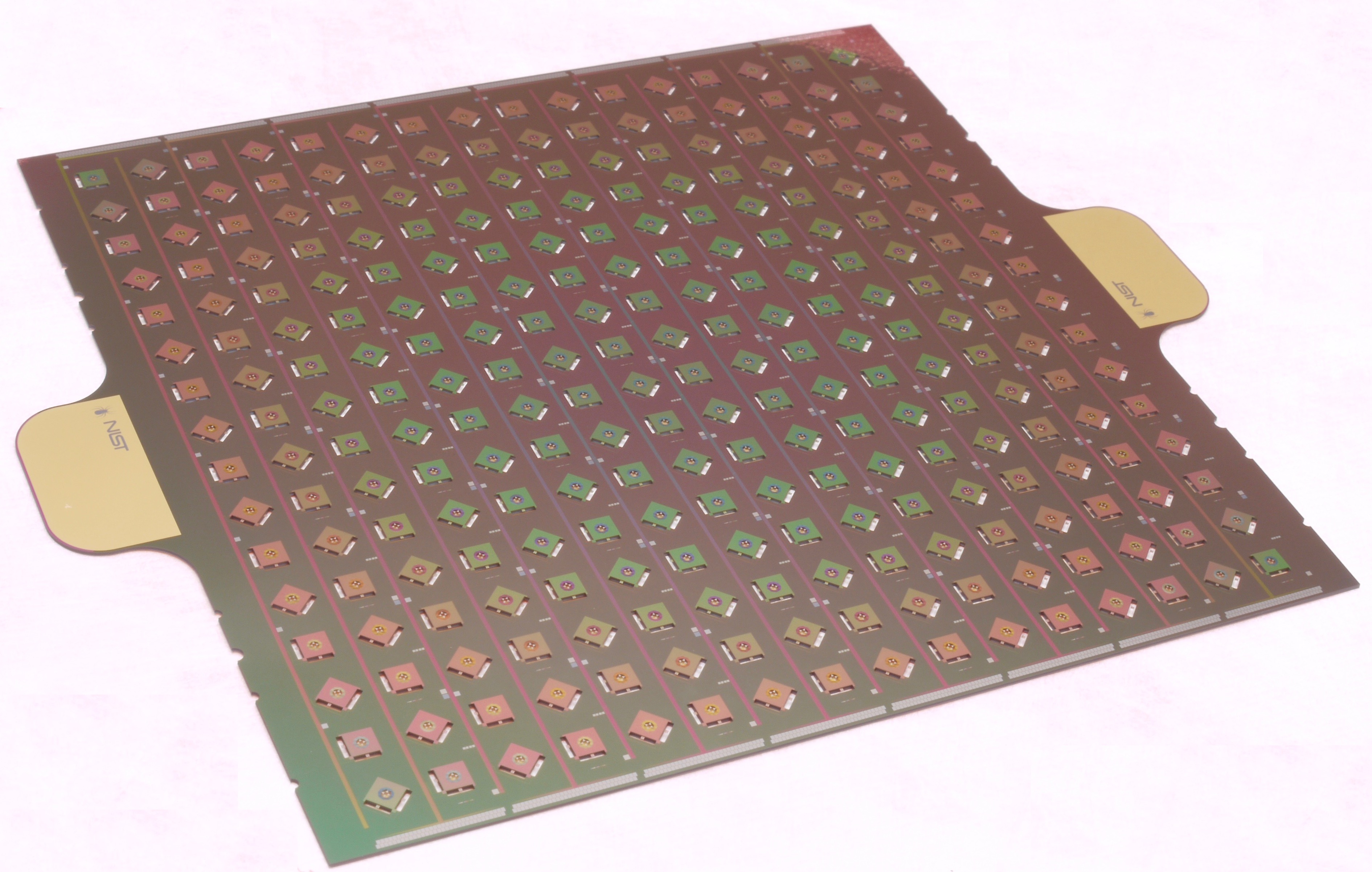}}
  \hspace*{.2in}
  \raisebox{-0.5\height}{\includegraphics[width=0.45\linewidth, keepaspectratio]{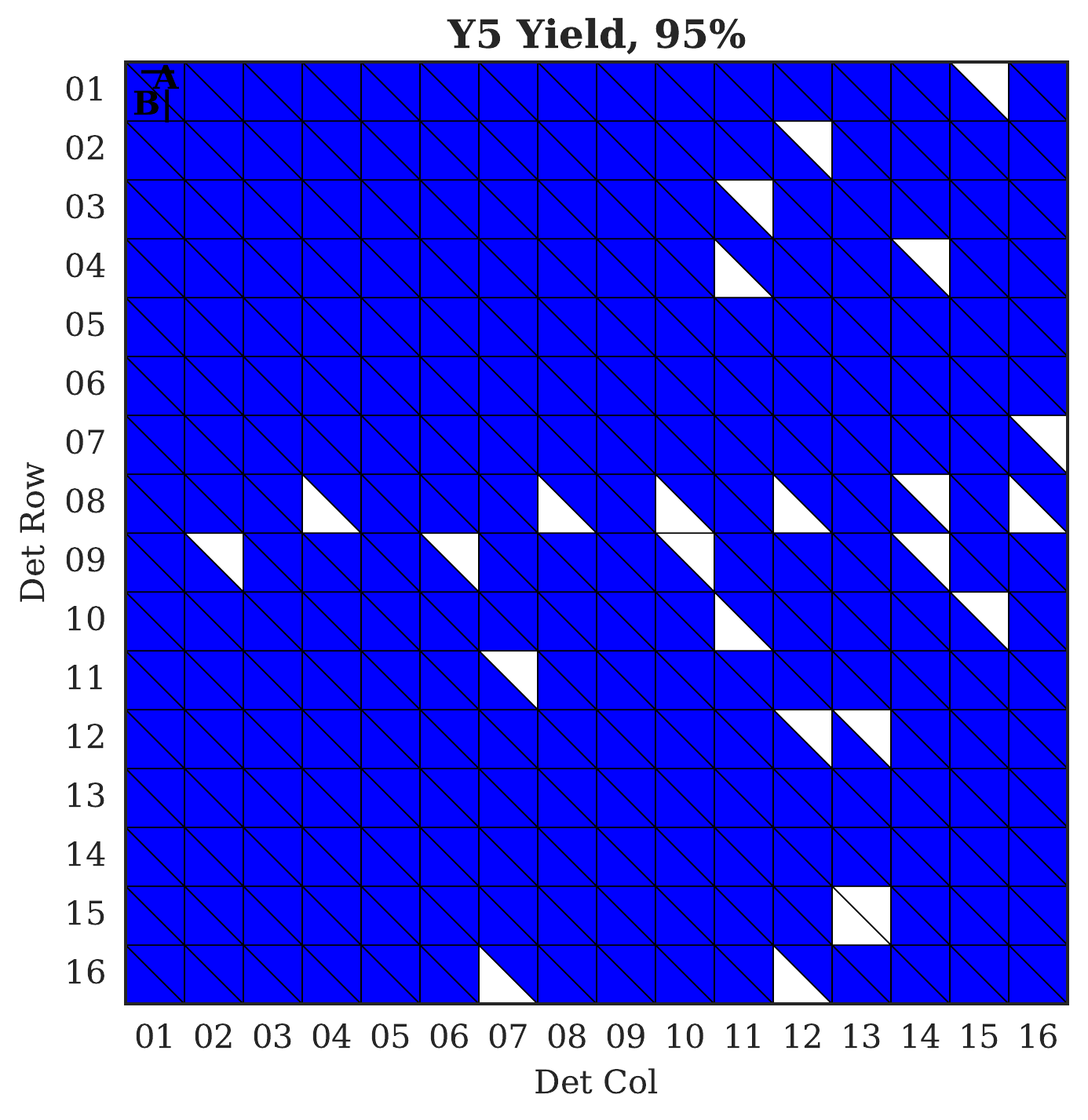}}
\end{minipage}
\caption{High-yield 150~mm wafer fabrication example.  
{\it Left:} Image of 512-TES bolometer array fabricated for the second flight of SPIDER.  
{\it Right:} A wafer map of functioning bolometers (colored blue) defined by acquiring a usable cryogenic I-V curve.   
Each square is divided in two to represent the X and Y polarization sensitive bolometers.    
The yield, which includes wiring and readout defects, is 95\%.  
See Bergman et al. \cite{bergman2017ltd} for details. (Color figure online.)
\label{fig:spideryield}}
\end{figure}

In order to deploy these projects efficiently, wafer throughput (defined as the product of rate and yield) of 150~mm arrays must be high.  
Historically, the yield of CMB detector arrays has been marginal. 
However, by streamlining the fabrication process of 150~mm CMB detector arrays as reported in Duff et al. \cite{duff2016advanced}, NIST has demonstrated the fabrication of deployment quality arrays in single fabrication runs.    
The Advanced ACTPol 150/220~GHz array \cite{ho2017highly}, Advanced ACTPol 90/150~GHz arrays \cite{choi2017characterization}, and SPIDER 280~GHz arrays \cite{bergman2017ltd} show high device yield, exhibit $<10\%~rms$ spread in bolometer saturation power, 
and each deliverable was produced in one fabrication run that used a wafer lot size between 2 and 4.     
\figref{fig:spideryield} illustrates the success of this fabrication process through the SPIDER 280~GHz detector array fabrication.  

In summary, we argue that the near-term community-wide production rates in combination with several recent demonstrations of high quality, 
high yield wafer production put the community in a strong position to address the sensor needs of upcoming ambitious experiments, such as CMB-Stage-IV.

\section{Conclusions}
The wave of excitement in CMB research has not yet crested as the CMB continues to be the gift that keeps on giving.  
Tremendous progress has been made in the development of superconducting ICs for CMB measurements in the past decade.  
The first generation ICs, which were single frequency band devices, have been used to make state-of-the-art measurements of the CMB angular power spectrum.  
These measurements have already provided new insights into cosmology and fundamental physics.
Newer generation ICs have expanded pixel bandwidth and several frequency bands per spatial pixel.  
These multichroic arrays have been deployed in multiple instruments and are soon to produce science results.

The success of next-generation experiments does not depend on the development of fundamentally new focal plane sensing technology, but in scaling up existing technology.  
Community-wide detector production capabilities of existing detector architectures are on-track to meet these demands.  
However, new instruments would greatly benefit from improvements in detector packaging, highly multiplexed readout, and high-throughput testing capability.  
Significant effort has already begun for all of these development areas.
Lastly, although new detector types are not essential to make progress in CMB observations, history has shown that enabling technologies quickly find application.      
In this regard, MKIDs optimized for CMB, hierarchical phased arrays, and other forward looking LTD technologies that make fielding CMB imagers easier and less expensive are efforts well spent.

\begin{acknowledgements}
We thank Lyman Page, Roger O'Brient, Adam Anderson, and Kam Arnold for figure contributions.   
\end{acknowledgements}
\pagebreak


\end{document}